\documentstyle[epsf]{mn}

\newif\ifAMStwofonts

\newcommand{\kms}{km~s$^{-1}$}
\newcommand{\cm}{cm$^{-2}$}
\newcommand{\lya}{Lyman~$\alpha$}

\newcommand{\nhi}{$N$(H~I)}

\def\ltsima{$\; \buildrel < \over \sim \;$}
\def\simlt{\lower.5ex\hbox{\ltsima}}
\def\gtsima{$\; \buildrel > \over \sim \;$}
\def\simgt{\lower.5ex\hbox{\gtsima}}

\ifoldfss
  \ifCUPmtlplainloaded \else
    \NewTextAlphabet{textbfit} {cmbxti10} {}
    \NewTextAlphabet{textbfss} {cmssbx10} {}
    \NewMathAlphabet{mathbfit} {cmbxti10} {} 
    \NewMathAlphabet{mathbfss} {cmssbx10} {} 
  \fi
  \ifAMStwofonts
    \ifCUPmtlplainloaded \else
      \NewSymbolFont{upmath} {eurm10}
      \NewSymbolFont{AMSa} {msam10}
      \NewMathSymbol{\upi}     {0}{upmath}{19}
      \NewMathSymbol{\umu}     {0}{upmath}{16}
      \NewMathSymbol{\upartial}{0}{upmath}{40}
      \NewMathSymbol{\leqslant}{3}{AMSa}{36}
      \NewMathSymbol{\geqslant}{3}{AMSa}{3E}

       \let\le=\leqslant
       \let\ge=\geqslant
    \fi
  \fi
\fi 

\ifnfssone
  \newmathalphabet{\mathit}
  \addtoversion{normal}{\mathit}{cmr}{m}{it}
  \addtoversion{bold}{\mathit}{cmr}{bx}{it}
  \newmathalphabet{\mathbfit} 
  \addtoversion{normal}{\mathbfit}{cmr}{bx}{it}
  \addtoversion{bold}{\mathbfit}{cmr}{bx}{it}
  \newmathalphabet{\mathbfss} 
  \addtoversion{normal}{\mathbfss}{cmss}{bx}{n}
  \addtoversion{bold}{\mathbfss}{cmss}{bx}{n}
  \ifAMStwofonts
    \ifCUPmtlplainloaded \else
      \UseAMStwoboldmath
      \makeatletter
      \new@mathgroup\upmath@group
      \define@mathgroup\mv@normal\upmath@group{eur}{m}{n}
      \define@mathgroup\mv@bold\upmath@group{eur}{b}{n}
      \edef\UPM{\hexnumber\upmath@group}
      \new@mathgroup\amsa@group
      \define@mathgroup\mv@normal\amsa@group{msa}{m}{n}
      \define@mathgroup\mv@bold\amsa@group{msa}{m}{n}
      \edef\AMSa{\hexnumber\amsa@group}
      \makeatother
      \mathchardef\upi="0\UPM19
      \mathchardef\umu="0\UPM16
      \mathchardef\upartial="0\UPM40
      \mathchardef\leqslant="3\AMSa36
      \mathchardef\geqslant="3\AMSa3E

       \let\le=\leqslant
       \let\ge=\geqslant
    \fi
  \fi
\fi 

\ifnfsstwo
  \DeclareMathAlphabet{\mathbfit}{OT1}{cmr}{bx}{it}
  \SetMathAlphabet\mathbfit{bold}{OT1}{cmr}{bx}{it}
  \DeclareMathAlphabet{\mathbfss}{OT1}{cmss}{bx}{n}
  \SetMathAlphabet\mathbfss{bold}{OT1}{cmss}{bx}{n}
  \ifAMStwofonts
    \ifCUPmtlplainloaded \else
      \DeclareSymbolFont{UPM}{U}{eur}{m}{n}
      \SetSymbolFont{UPM}{bold}{U}{eur}{b}{n}
      \DeclareSymbolFont{AMSa}{U}{msa}{m}{n}
      \DeclareMathSymbol{\upi}{0}{UPM}{"19}
      \DeclareMathSymbol{\umu}{0}{UPM}{"16}
      \DeclareMathSymbol{\upartial}{0}{UPM}{"40}
      \DeclareMathSymbol{\leqslant}{3}{AMSa}{"36}
      \DeclareMathSymbol{\geqslant}{3}{AMSa}{"3E}

       \let\le=\leqslant
       \let\ge=\geqslant
    \fi
  \fi
\fi 

\ifCUPmtlplainloaded \else
  \ifAMStwofonts \else 
    \def\upi{\pi}
    \def\umu{\mu}
    \def\upartial{\partial}
  \fi
\fi

\title{The First Detection of Cobalt In A Damped Lyman Alpha System}
\author[S. Ellison et al.]
       {S. L. Ellison$^1$, S. G. Ryan$^2$, J. X. Prochaska$^3$ \\
       $^1$European Southern Observatory, Casilla 19001, Santiago 19, 
	Chile\\
       $^2$Dept of Physics \& Astronomy, The Open University, Walton Hall,
	Milton Keynes, MK7 6AA, UK\\
       $^3$Carnegie Observatories, Pasadena, CA 91101, USA}
\date{}

\pagerange{\pageref{firstpage}--\pageref{lastpage}}
\pubyear{2001}

\begin{document}

\maketitle

\label{firstpage}

\begin{abstract}
The study of elemental abundances in Damped Lyman Alpha systems 
(DLAs) at high redshift represents one of our best opportunities 
to probe galaxy formation and chemical evolution at early times.  
By coupling measurements made in high $z$ DLAs with our knowledge of 
abundances determined locally and with nucleosynthetic models,
we can start to piece together the star formation histories of 
these galaxies.  Here, we discuss the clues to galactic chemical
evolution that may be gleaned from studying the abundance of Co
in DLAs.  We present high resolution echelle spectra of two QSOs, 
Q2206$-$199 and Q1223+17, both already known to exhibit intervening 
damped systems.  These observations have resulted in the first ever
detection of Co at high redshift, associated with the
$z_{abs}= 1.92$ DLA in the sightline towards Q2206$-$199.
We find that the
abundance of Co is approximately $\frac{1}{4} Z_{\odot}$
and that there is a clear overabundance relative to iron, 
[Co/Fe] = $+0.31\pm0.05$.  From the abundance of Zn, we determine that
this is a relatively metal-rich DLA, with a metallicity 
approximately $\frac{1}{3} Z_{\odot}$.  Therefore, this first
detection of Co is similar to the marked overabundance relative to Fe seen
in Galactic bulge and thick disk stars.   
\end{abstract}


\section{Introduction}
Our understanding of the processes associated with galactic chemical
enrichment, such as stellar nucleosynthesis, dust depletion and 
mixing mechanisms, can benefit greatly from the comparison of
local abundances with those measured at high redshift.  At low $z$,
abundances can be determined for the Milky Way and its near 
neighbours from stars, H~II regions and the interstellar 
medium (e.g. Chen et al. 1999; Kobulnicky and Zaritsky 1999;
Shetrone et al. 2000; Izotov and Thuan 1999; Savage and Sembach 1996).
At high redshift, the most accurate and extensive abundance measurements
are those determined from Damped Lyman Alpha systems 
(DLAs).  These systems are the highest H~I column density
members of the quasar absorption line menagerie and are traditionally
defined as having N(H~I) $\ge 2 \times 10 ^{20}$ atoms \cm. 
Coupling measurements of metal abundances in high redshift DLAs with
our knowledge of local enrichment patterns plays a crucial role
in our understanding of the star formation histories of these distant
objects and the early stages in galaxy evolution.

In the context of investigating galactic chemical evolution,
several key elements have previously been targeted as particularly
informative diagnostics.  The utility of these key metals is often based on 
our knowledge of their relative production and release timescales 
in stars (and subsequent supernovae) of different masses.
For example, the $\alpha$ elements (Mg, Si, Ca etc.)
have been extensively used to
investigate the amount and rate of massive star formation.  An overabundance
of $\alpha$ relative to Fe in metal-poor Galactic halo  
stars chronicles the epoch of early star formation (Tinsley 1979; Ryan, Norris
and Beers 1996) and
in the bulge the same overabundances in solar metallicity stars
are indicative of a `get rich quick' population (McWilliam and
Rich 1994; Rich and McWilliam 2000) i.e. one which is as old as 
the Galactic halo but as metal-rich 
as the Galactic disk.  In DLAs, there is on-going discussion
about the abundances of $\alpha$ elements, for which nucleosynthetic origins
of abundance patterns must be disentangled from dust depletion effects.  
Although there is some evidence that Si may be overabundant in DLAs
(Lu, Sargent and Barlow 1998), [Si/Zn] is solar in at least some systems
(Pettini et al 2000).  Moreover, since Si is highly refractory,
the interpretation of [Si/Zn] measurements relies on a dust 
correction factor based on [Zn/Cr], which is quite uncertain.
Conversely, the abundance of S, an $\alpha$ element not depleted
onto dust, has been found by Centurion et al. (2000) to be
underabundant relative to Fe, a result at odds with the Type II supernova
pattern observed in the Milky Way. These results have
been interpreted as possible evidence for star formation having 
occurred at a low rate or in sporadic bursts.  However, the Centurion 
et al. (2000) data were obtained with resolution R = 5000
and need to be followed up with higher resolution data in order reliabley
distinguish the S lines from the \lya\ forest.
Further clues to the star formation history of DLAs
may be gleaned from N/O ratios.  The large observed spread in N/O 
in DLAs can be explained by the delayed release of primary
N (relative to O) in metal-poor intermediate mass stars 
(Pettini et al. 2001; Lattanzio et al. 2001).

Here we focus on cobalt, an element which although relatively
widely studied in Galactic stars, has not yet been featured in
DLA abundance studies.  The intriguing trends of [Co/Fe]
as a function of [Fe/H] in the various Galactic stellar
populations indicate that cobalt may be another element
that can provide important clues to chemical evolution.  
In sub-solar metallicity thin disk stars, Co is found to be slightly
underabundant with respect to Fe ([Co/Fe] $\sim -0.1$, 
Gratton and Sneden 1991), although this may be related to their solar 
log~$gf$ scale being on average 0.06$\pm$0.04 higher than the 
laboratory scale adopted by
many other observers (see Section 5.1 for more details).   This trend
continues in moderately metal-poor halo stars down to metallicities of
[Fe/H] $\sim -2.5$.  For more metal-poor halo stars, a considerable
overabundance of Co with respect to Fe is observed, $0 <$ [Co/Fe] $< 0.8$,
(McWilliam et al. 1995; Ryan, Norris and Beers 1996).  Interestingly, this
coincides with a downturn in relative Cr to Fe abundances, such that Co and
Cr are closely anti-correlated in the metal poor population. 
Similar [Co/Fe] overabundances are observed in the bulge 
([Co/Fe] $\sim +0.3$, McWilliam \& Rich 1994), which, we recall, 
is as old as the halo, and to a lesser degree in the
thick disk ([Co/Fe] $\sim +0.1$, Prochaska et al. 2000).

The detection of Co~II transitions is a challenging possibility for
ISM absorption line spectroscopy, both locally and at high $z$.  
Although there are numerous transitions
that lie at convenient rest wavelengths for both local and high
redshift studies, only two interstellar measurements currently exist.
These detections have been achieved in the local ISM in the sightlines
towards $\zeta$ Oph (Federman et al. 1993) and $\rho$ Oph A (Mullman 1998b)
using the Co~II $\lambda$1466 transition which has an equivalent width
in these interstellar clouds of only 0.5 and 4.7 m\AA\ respectively.  
One of the advantages of moving to higher redshifts is that the relatively
strong UV lines become shifted into the optical where one has
access to more efficient detectors and larger telescopes.  In addition,
the observed equivalent width is
increased by a factor of ($z + 1$).  However, the fundamental problem  
remains that the Co~II lines are all intrinsically weak and 
that the solar abundance of Co is $8 \times 10^{-8}$ that
of hydrogen.  In addition,
Mullman et al. (1998a,b) have provided new laboratory measurements for
Co~II $f$-values (oscillator strengths), typically revising the old
values downwards by a factor of approximately two.  This means that
the spectral detection limit has effectively been increased by the
same factor.  Finally, 
the two local detections have brought to light an additional
problem, namely that cobalt appears to be a highly refractory element
that is depleted onto dust to a similarly high degree as iron (Mullman
et al. 1998b).  

Despite the technical challenge of observing Co in the high redshift
ISM of DLAs, it is now feasible with high resolution spectrographs
on 8 -- 10 m class telescopes to detect this element, in reasonable
exposure times, down to log $N$(Co) $\sim$ 12.5.   
Here we present the first ever detection of 
Co in a DLA, at $z_{\rm abs} = 1.92$, determined from a spectrum obtained with
the Ultra-violet and Visual Echelle
Spectrograph (UVES) on UT2 of the VLT.  In addition, we present a combined UVES
and HIRES (High Resolution
Echelle Spectrograph) spectrum obtained at the VLT and Keck telescopes 
respectively, which can also offer an interesting upper
limit to [Co/H] in a second DLA.  We discuss the implications for
the star formation history of these high redshift galaxies and
compare our observations with recent nucleosynthetic yield models.

\section{Target Selection}

In order to at least place interesting limits on the abundance of
Co in DLAs, it is desirable to reach a sensitivity level of 
[Co/Fe] = 0.  Such a detection limit is sufficient at least to
distinguish between the marked overabundance of Co relative to Fe seen in the
halo, bulge, and to a lesser degree in the thick disk, and the mild
underabundance in the Galactic thin disk.  Therefore,
in order to maximise the likelihood of a Co~II detection, we have selected
bright QSOs with foreground DLAs known to contain high column densities
of Fe.  This criterion can either be achieved in moderate metallicity
DLAs with high \nhi\ or in lower column density systems with relatively 
high levels of enrichment.  The former type of system is relatively rare
due to the power law distribution of column densities in H~I absorbers
which shows that systems with \nhi\ $\simgt$ 2 $\times 10^{21}$ \cm\ are
very scarce.  Similarly, few metal rich ($Z > \frac{1}{10} Z_{\odot}$)
DLAs have been identified at any redshift (Pettini et al. 1999),
indicating that extinction due to dust may be biasing current samples
(Ellison et al. 2000).  Finally, as dust extinction may be greater 
in metal-rich and high-$N$(H) DLAs (e.g. Prantzos and Boissier 2000),
it is particularly challenging to identify QSOs absorbed by such systems
that also are bright enough to provide the requisite S/N.

Having searched the literature for candidate targets, we 
initially identified two QSOs for study, Q2206$-$199 ($B = 17.8,
z_{em} = 2.559$) and Q1223+17 ($B = 18.5, z_{em} = 2.936$).
The former of these targets actually exhibits 3 separate DLAs
in its absorption spectrum (Prochaska and Wolfe 1997); here
we focus on the system at $z_{abs} = 1.92$.  Although this system
only has a moderate column density compared with other DLAs
(\nhi\ = 4.8$\times 10^{20}$ \cm, see section 4), it is amongst the most
metal-rich systems known with $Z \sim \frac{1}{3}Z_{\odot}$.  
In fact, Prochaska
and Wolfe (1997) were able to detect some elements rarely seen
in DLAs in this system, such as Ti.  Conversely, Q1223+17
is a very high column density absorber, \nhi\ = 3$\times 10^{21}$ \cm\
and therefore has a relatively high column density of Fe despite its low
metallicity (Pettini et al. 1994; Prochaska et al. 2001).

\section{Observations}

We present high resolution echelle spectra for 
these two high redshift QSOs.
The data were obtained with UVES on the VLT (Q2206$-$199 and Q1223+17) 
and with HIRES on the Keck telescope (Q1223+17).
A summary of the observations is given in Table~1.

\begin{table*}
\begin{center}
\caption{Summary of Observations}
\begin{tabular}{llllcr} \hline \hline
Quasar &Telescope/Instrument & 
Max. $\lambda$ Coverage & Resolution &Int Time (s)  & S/N \\ \hline
Q2206$-$199  &  VLT/UVES  & 3300 -- 6650 \AA & 42\,000 & 17\,100 & 20 -- 70 \\ 
Q1223+17  &  VLT/UVES & 3800 -- 10250 \AA & 45\,000 & 10\,800  & 20 -- 50 \\ 
Q1223+17 & Keck/HIRES & 4750 -- 7200 \AA & 38\,000 & 19\,100 & $\sim 30$ \\ \hline
\end{tabular}
\end{center}
\end{table*}

\subsection{Details of Observations and Data Reduction of Q2206$-$199}

Spectra of Q2206$-$199 were obtained over 3 nights from May 28-30 2000
with UVES on the VLT
in conditions of good seeing, typically 0.5 -- 0.7 arcseconds,
although for brief periods extremes of 0.4 and 1.3 arcsecs were reached.
A slit width of 1.0 arcsec
was fixed throughout the observations, oriented along the parallactic angle.  
The data presented here were obtained with the dichroic 1 $\lambda_{cen}
= 390+564$ 
configuration and read out in 2 $\times$ 2 pixel-binned, high-gain mode.
The blue arm CCD is a single thinned EEV chip whereas the red arm 
consists of a mosaic of a second EEV chip and an MIT chip, separated
by approximately 0.96 mm.   A single dichroic setting therefore
has non-contiguous wavelength coverage and in
the $\lambda_{cen} = 390+564$ setting there are gaps between 
4525 -- 4625 and 5600 -- 5680 \AA. 

The extraction of the spectra was achieved using a customized version
of the UVES pipeline which is based on ECHELLE routines in
the data reduction package MIDAS.  A detailed description of this
process can be found in Ballester et al. (2000).  
The spectra were optimally extracted separately for the three CCDs and the
variation in the continuum flux level, due primarily to the blaze function of
the echelle grating, was removed.
The resolution of the spectra, as determined from ThAr arc lines, is
$R \sim 43\,000$, or approximately 7 \kms\ FWHM.  1-D spectra were
obtained by joining the orders and converting the wavelengths to a
vacuum heliocentric scale.  
The individual spectra were then co-added with a weighting proportional
to their S/N.  The final step was to normalise the spectrum by dividing
through by a spline function fitted through absorption-free regions
of data.   Since most of the metal lines which we study here are
found redward of the \lya\ emission, the continuum can be accurately
fitted since there are relatively few absorption features and the
S/N is quite high.

\subsection{Details of Observations and Data Reduction of Q1223+17}

The spectrum presented here of the $z_{abs} = 2.466$ DLA towards
Q1223+17 is a composite of HIRES plus UVES data.  The UVES
data were obtained during the same run as the Q2206$-$199 spectra
and reduced in an identical way.  The only difference for the
data of Q1223+17 is that they were obtained with the dichroic 2 $\lambda_{cen}
= 437+860$ setting.
This mode offers a larger, redder overall wavelength coverage at the expense
of slightly
larger inter-arm gaps (between 5000 -- 6750 \AA\ and 8500 -- 8750 \AA).
In addition to the UVES data, we have obtained $\sim$ 5 hours
 of Keck HIRES data, which are described in more detail by the 
comprehensive abundance analysis of Prochaska et al. (2001).  In brief,   
the C5 decker was implemented providing
FWHM~$\approx 8$ \kms\ resolution with the setup covering
$\lambda \approx 4750 - 7200$~\AA.  The data were reduced with the 
MAKEE software package developed by T. Barlow
for the extraction of HIRES spectra.  
The coadded spectrum has S/N~$\approx 30$ per 2 \kms\ pixel.

To obtain the strictest upper limit on the Co~II column density for
the damped \lya\ system towards Q1223+17, we coadded the HIRES and UVES
spectra in the region spanning the Co~II $\lambda$2012 transition.  
Each spectrum was 
rebinned to a common wavelength scale and then optimally coadded resulting
in a final spectrum with S/N$~\approx 36$ per 2 \kms\ pixel.

\section{Neutral Hydrogen And Cobalt Column Densities}
 
\begin{figure}
\epsfxsize=084mm
\epsfbox[44 045 584 772]{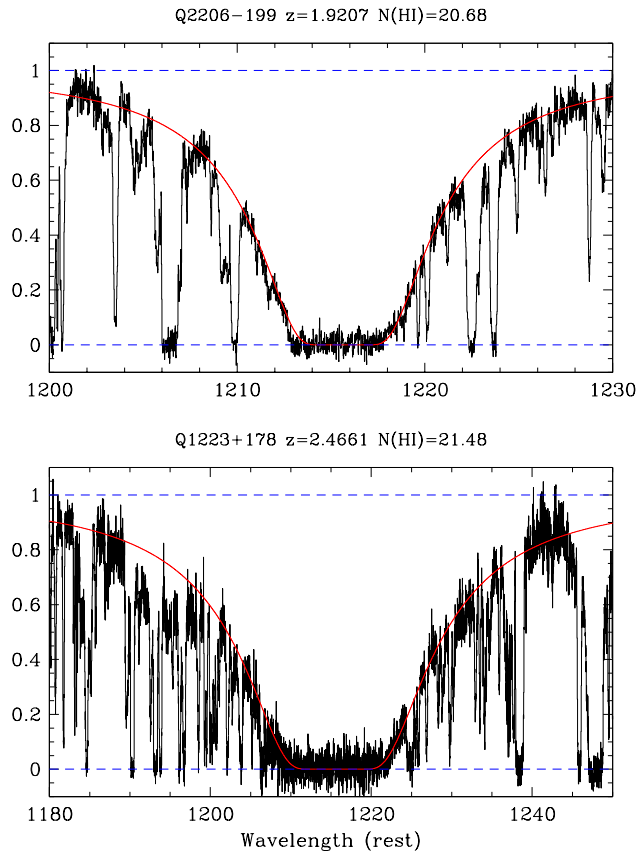}
\caption{Fits to the damped \lya\ lines for the two QSOs studied here.
The data are shown in the rest wavelength for the absorption redshift
stated in each panel. }
\end{figure}

The H~I column density of the two QSOs was determined
by fitting fully damped profiles to the normalised spectra using
the Starlink package Dipso.  The UVES spectrum of Q2206$-$199 was found
to have a best fit \nhi\ = 4.8$\pm 0.3 \times 10^{20}$ \cm\ and
a redshift of $z_{abs} = 1.9207$, Figure~1. 
Note that this fit is very well constrained by the shape of the central 
trough and damping wings 
and the \nhi\ is in good agreement with previously determined estimates 
(e.g. Pettini et al. 1994) for this DLA.   The HIRES spectrum of Q1223+17
exhibits a relatively large column density of H~I which is found to
have a best fit of \nhi\ = 3$\pm 0.3 \times 10^{21}$ \cm\ and
a redshift of $z_{abs} = 2.4661$, also in good agreement with the
Pettini et al. (1994) measurement.

A detailed analysis of other metal abundances in the DLAs
towards Q2206$-$199 and Q1223+17 can be found in Prochaska and Wolfe
(1997) and Prochaska et al. (2001); here we simply concentrate on
the detection of Co~II.  However, the analysis of other metal lines
plays an important role in determining whether lines are blended
and distinguishing weak features from the noise.  One of the standard 
procedures is to fit Voigt profiles to the absorption
system which is described as a complex of `clouds', each defined by
a Doppler parameter ($b$-value), column density ($N$) and redshift.
It has been unanimously found in other high resolution DLA studies that
singly ionized species such as Fe~II, Si~II and Ni~II can be fitted
using identical cloud models (e.g. Pettini et al 1999).  
Therefore, line-fitting is often
performed first for relatively strong lines with several observed 
transitions, such as Fe~II, in order to fix the $b$-values and redshifts
of the constituent components.  For the other species, excellent
profile fits can be achieved by allowing only the column density to
vary as a free parameter, indicating that the singly ionized metal 
lines have similar kinematical structure.  Fixing the cloud model
therefore 
offers a significant advantage when fitting weak lines, especially if
there is a risk of some contaminating absorption from other species
at a different redshift, e.g. contamination in the Ly$\alpha$ forest.

\subsection{Detection Of Cobalt in the DLA Towards Q2206$-$199}

In the case of Q2206$-$199, the absorption system is quite complex
in comparison with other DLA metal line systems.  The fit was initially
constrained using Fe~II $\lambda$2249 and Fe~II $\lambda$2260 and
refined by performing a simultaneous fit with Ni~II $\lambda$1750,
Ni~II $\lambda$1741 and Ni~II $\lambda$1709.  Together, these
five transitions can provide excellent constraints on the cloud
model and, in addition, are two elements that when considered
along with Co may provide interesting clues to nucleosynthetic processes
(see section 5).  The $b$-values and redshifts
of each cloud were tied within the fitting procedure such that 
these parameters could vary, but remained identical for the two
elements.  The best fit results for this procedure yielded a 6 component
cloud model which extends over almost 150 \kms, see Table 2.
Having determined the cloud model for Fe~II and Ni~II, we
use this template to fit Co~II (and the other metal lines), 
again allowing only the column densities to vary.

\begin{table*}
\caption{Voigt Profile Fit Parameters for Fe-Peak Elements in the 
DLA Towards Q2206$-$19. }
\begin{tabular}{cccccccc}\hline \hline
Cloud &Redshift & $b$& 
\multicolumn{5}{c}{Log$_{10}$ N(X)}\\ && &Co~II&Ni~II&Fe~II&
Zn~II&Cr~II\\ \hline
1 &  1.91995 &   14.8 & 11.54 &   13.09 & 14.42 & 11.88 & 12.56\\
2 &  1.92000 &    5.9 & 12.51 &   13.53 & 14.58 & 12.10 & 12.88\\
3 &  1.92023 &   11.0 & 11.81 &   13.17 & 14.22 & 11.90 & 12.42\\
4 &  1.92040 &    5.1 & 12.05 &   12.95 & 14.15 & 11.87 & 12.38\\
5 &  1.92062 &   14.9 & 12.45 &   13.70 & 14.87 & 12.45 & 13.17\\
6 &  1.92094 &   31.9 & 12.61 &   13.69 & 14.79 & 12.41 & 13.06\\ \hline
\multicolumn{3}{c}{Total system N(X)} & 13.09$\pm0.05$ & 14.23$\pm0.01$ &
15.36$\pm0.01$ & 12.95$\pm0.01$ & 13.63$\pm0.01$ \\ \hline
\end{tabular}
\end{table*}

The revised $f$-values of Co (Mullman et al. 1998a,b) show that the
two strongest ground state resonance lines covered by this spectrum
are at rest wavelengths of 1466.21 \AA\ ($f$=0.031) and 2012.16 \AA\
($f$=0.037).  A third line of similar strength is located at
$\lambda$ = 1941.29 \AA\ ($f$=0.034) but falls in the gap between the
two red arm CCDs.  The line at 1466 \AA\ for a system at $z_{abs}=1.92$
lies in the \lya\ forest of the $z_{em}= 2.559$ QSO.  
The Co~II $\lambda$2012 line lies redward 
of the QSO's \lya\ emission where the risk of mis-identification or
contamination is much lower.  

\begin{figure}
\epsfxsize=084mm
\epsfbox[20 040 585 761]{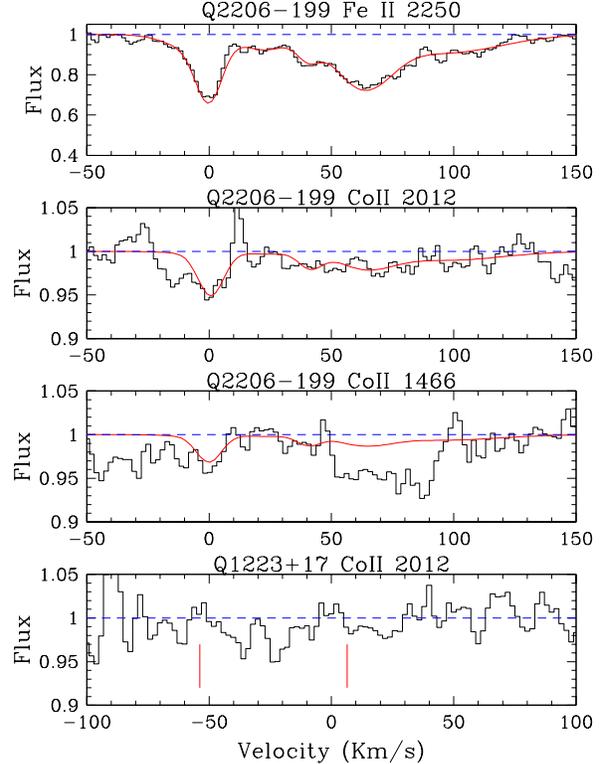}
\caption{Fits to Fe~II $\lambda$2250 and the two Co~II lines observed 
towards Q2206$-$199
and the non-detection towards Q1223+17.  In all four panels,
the data are shown by the solid histogram and the continuum level 
by the dashed line.  For Q2206$-$199, the fit is shown by the solid curve,
and the vertical tick marks in the bottom panel mark the expected 
positions of the two strongest components in Q1223+17.  The Q2206$-$199 fit 
was achieved by using Fe~II and Ni~II transitions to constrain the
cloud model and then allowing $N$(Co) to vary as a free parameter.  The
Co~II $\lambda$1466 line is blended in the \lya\ forest and is not used
to improve the final estimate of $N$(Co). }
\end{figure}

The $\lambda$2012 line was used to 
determine the $N$(Co) of individual components given
in Table 2 which also lists the fits to the other Fe-peak elements
covered by our spectrum.  Although the column density error for any individual 
component can be quite significant, particularly for closely-spaced, blended
lines, the error on the total N(X) is usually substantially
smaller.  Also, the error clearly depends on the wavelength of the
transition (since the S/N varies along the spectrum) and is much 
reduced if several transitions are used in the fit.  
All of these factors are taken into account by VPFIT when determining
the total column density errors from the chi-squared reduction.
In addition, we consider the effect of the continuum fit on our results,
which may be particularly important for Co
since a large fraction of the absorption is associated with
relatively broad, shallow features.  
We firstly note that the continuum fit was done
with a low order polynomial across the flat continuum around the
CoII $\lambda$2012 line and that any significant departure from this 
would require quite a contrived placement.  Secondly, VPFIT may
adjust the continuum in order to improve the fit, although
no change was made in this case.  As can be seen in Table 3, the
relative amounts of Co in the red and blue components are in good
agreement with the other elements, a sign
that at least the shape of the continuum is correct.
In terms of a systematic offset, in Figure 2 it can be seen that 
had the continuum been placed too low, there would be clear evidence 
of flux above the unity level.  If the continuum had been placed
too high, this would only increase $N$(Co), thus reinforcing the
observed overabundance of [Co/Fe].  Nonetheless, we have investigated 
the possible
systematic uncertainty induced by continuum placement around Co~II
and other elements and determine the effect on derived column densities
to be not more than 0.10 dex, and typically significantly less than
this.  

Over-plotting the fit from Co~II $\lambda$2012 onto the Co~II $\lambda$1466
line (Figure 2) shows that although there is evidence for weak 
Co~II absorption there is considerable blending, probably due to \lya\
forest lines.  There is also some possibility of blending
in the $\lambda$2012 transition, since it can be seen in Figure~2
that an additional component is present at the blue edge of
the absorption complex that is not fitted by the cloud model
we adopt (see the top panel of Figure 2 for a comparison with the
Fe~II $\lambda$2250 line).  In order to investigate the possible impact 
of this component on our column density determination, we consider the
relative contributions of the individual clouds to the total
absorption seen in the other metal lines.  Table 3 shows the 
column density of metal enriched gas in the blue ($N(X)_b$) and
red ($N(X)_r$) components, where the blue component is defined to
be made up of clouds 1 and 2 from Table 2, whereas the red component
is taken to consist of  clouds 3 -- 6.  Also listed in Table 3 is
the fraction of gas in the blue component ($f_b$)
for all of the metals measured in this DLA.  From these statistics
it can be seen that, excluding Co, on average 27\% of the enriched gas
resides in the blue component, a value which remains remarkably constant for
all the ions considered here. For Co, however, we see that a slightly
higher fraction of gas, 29\%, resides in the blue component.
This small increase may be caused by the wing of the contaminating
line.  We attempted to identify the possible cause of contamination,
but were unable to associate the extra component with any metal
line from either of the other 2 DLAs.  
To show that the effect of the blending is probably negligible,
we assume that the fractional distribution
of gas is the same for Co as for the other metals and that therefore
the $N_r$(Co)=12.94 represents 73\% of the total, implying a total
$N$(Co) = 13.08, almost exactly what is determined from the original fit.

\begin{table}
\begin{center}
\caption{Column Densities in the Blue and Red Components of the DLA
Towards Q2206$-$199}
\begin{tabular}{cccc} \hline \hline
Element, $X$ & Log $N(X)_b$& Log $N(X)_r$ & $f_b$ \\ \hline
Co & 12.55 & 12.94 & 0.29 \\
Fe & 14.81 & 15.22 & 0.28 \\
Ni & 13.67 & 14.09 & 0.28 \\
Zn & 12.31 & 12.84 & 0.23 \\
Cr & 13.05 & 13.50 & 0.26 \\
Si & 15.23 & 15.63 & 0.28 \\ \hline
\end{tabular}
\end{center}
\end{table}

\subsection{The Upper Limit of Cobalt in the DLA Towards Q1223+17}

There is no Co~II $\lambda$2012 absorption visible in the combined
UVES+HIRES spectrum of Q1223+17, see Figure 2.  To place an upper limit on
the $N$(Co~II) value and in turn the Co/Fe ratio, we integrated the
apparent column density\footnote{$N_a(v) = m_e c \tau_a(v)/\pi e^2 f \lambda$
where $\tau_a(v) = \ln [I_i(v)/I_a(v)]$, $I_i$ and $I_a$ are the incident
and observed intensity, $f$ is the oscillator strength, and 
$\lambda$ is the rest wavelength of the transition (Savage \& Sembach 1991).}
over a small region covering the Co~II $\lambda$2012 transition.  
For high S/N data
(i.e.\ in the limit that $\ln (1-x) \approx x$), this is equivalent
to deriving the equivalent width over the same velocity range and then
calculating the column density assuming the linear curve of growth.
Although the stronger metal line transitions span over 200 \kms\
(e.g.\ Fe~II 1608), to place the tightest constraints on the Co/Fe ratio
we integrated from $-20$~km/s to $+20$~km/s which covers the strongest
Fe~II feature in the velocity profile.  Restricting our analysis to
this velocity range, we find $N$(Co~II)$ < 12.63$,
$N$(Fe~II)$ = 15.05$ and [Co/Fe]~$< +0.18$ where the limits
are at the $3 \sigma$ level.

\subsection{Corrections Due to Ionization and Dust?}

It is usually assumed in DLA abundance studies that the column densities
of the singly ionized metal species are a good approximation to the
total column density of this element, based on observations and theoretical
modeling (e.g. Viegas 1995).  However, there has been some
suggestion (Howk and Sembach 1999; Izotov et al. 2000) that ionization
corrections may in fact be important when deriving DLA abundances.
Although there is some temptation to invoke ionization corrections
to explain a few unusual abundance trends in DLAs, the
predictions of these models are often not borne out by 
observations (Pettini et al. 2000; Lattanzio et al 2001). 
In addition, Levshakov et al. (2000) have recently
shown that the physical conditions adopted by these models
are unrealistic when compared with those observed in local
DLA analogs.  Here, we are specifically
concerned with the relative abundances of Co and Fe.  Even in the
unlikely event that some correction is relevant to the absolute 
abundances as suggested by Izotov et al. (2000), the first and second 
ionization stages of cobalt and iron have very similar potentials 
(e.g. 7.88 eV and 7.90 eV for Co~I and Fe~I)
so that it is highly unlikely that an ionization correction is required
when converting N(Co~II) and N(Fe~II) to [Co/Fe].  The same is
true for nickel, since the ionization potential for Ni~I is 7.64~eV.   

The presence of dust in DLAs and its tendency to remove a fraction
of the total metals from the gas phase has now been well documented
(e.g. Pettini et al. 1994; Pettini et al. 1997).  This renders the
task of determining the true abundances in these systems somewhat
uncertain, although several attempts have been made to recover intrinsic
abundances by correcting for Galactic depletion patterns (Vladilo 1998;
Savaglio 2000).  In the case of Co, some of the more interesting
Galactic (stellar) abundance patterns are seen not in absolute (e.g. [Co/H])
measurements, but in the values of relative abundances like [Co/Fe]
and [Co/Cr].  The important consideration is therefore the
\textit{relative} depletion fractions of these elements, which can
be determined from local interstellar observations.  From Savage
and Sembach (1996) and Mullman et al. (1998b) we can see that 
Co, Cr, Ni and Fe are all depleted to a very similar level.  In addition,
at relatively low levels of depletion, as is the case for Q2206$-$199,
the small differences between the dust correction required for these
elements becomes negligible (Pettini et al. 2000).

Therefore, we assume that we can safely investigate relative abundance
ratios such as [Co/Fe] without the need for either corrections
for dust or ionization.

\section{Discussion Of Cobalt Abundances}

In Table 4, we list the abundances of the iron peak elements
(Fe, Ni, Cr, Zn, Co) measured for the two DLAs analysed herein and
two other systems found in the literature for which useful [Co/H]
upper limits can be determined.  

\begin{table*}
\begin{center}
\caption{Iron-peak abundances.  
Co limits are 3$\sigma$.  Note that the Ni abundance for 
HE1104$-$1805 has been increased by a factor of 1.9 from the published value, 
based on the revised $f$-values of Fedchak and Lawler (1999)}
\begin{tabular}{lcccccl} \hline \hline
Quasar & [Fe/H]&[Ni/H]& [Co/H]& [Cr/H]& [Zn/H] & Ref \\ \hline 
Q2206$-$199  & $-$0.81 & $-$0.70 & $-$0.50 & $-$0.73 & $-$0.42 &This work \\ 
Q1223+17  & $-$1.94 & $-$1.80  & $< -1.76$ & $-$1.66 & $-$1.94 & 
Prochaska et al. (2001), this work\\ 
Q0302$-$223 &$-$1.20 &$-$1.04& $< -0.87$ & $-$0.96 & $-$0.56&
Pettini et al. (2000) \\ 
HE1104$-$1805 & $-$1.59 & $-$1.40 & $< -1.56$  & $-$1.46 & $-$1.02&  
Lopez et al. (1999)\\ \hline
\end{tabular}
\end{center}
\end{table*}

\subsection{Homogeneity of Stellar Cobalt Abundances}

The cobalt abundances measured in the four DLAs will be compared 
in Section 5.2 with those for stars spanning a range of Galactic
populations and metallicities
(Sneden \& Gratton 1991; McWilliam \& Rich 1994;
McWilliam et al. 1995; Ryan, Norris, \& Beers 1996;
Prochaska et al. 2000).
The stellar measurements use solely Co~I lines. 

All stellar measurements
utilised laboratory $gf$ values from Cardon et al. (1982), with the 
exception of Sneden \& Gratton who derived solar $gf$ values.
These two scales appear to be very similar; five lines in common give a mean 
difference log~$gf_{\rm GS90} -$ log~$gf_{\rm CSSTW82}$ = +0.06~dex with a 
standard error in the mean of 0.04~dex. As this is significant only 
at the 1.5$\sigma$ level, it would be premature to
adjust for it. Furthermore, by using {\it solar} $gf$ values, Sneden and 
Gratton have effectively analysed their stars differentially with respect to 
the Sun, so changing the $gf$ values would alter the inferred solar abundance 
as well. We note, nevertheless, that even in their most metal-rich stars, 
[Co/Fe] tends towards $\simeq -0.05$ at [Fe/H] = 0, which supports the view 
that their Co abundances are marginally low. The problem (assuming there is 
one) may perhaps be traced to differences in the assumptions in the solar 
model used for their solar analysis and those employed in the other stellar
calculations. We did, however, correct 
the three stars of Sneden \& Gratton affected by the probable 0.47~dex error 
in the solar log~$gf$ value of the 4118~\AA\ line; see 
Norris, Ryan \& Beers (2001) for details.
A comparison with the 
improved $gf$ values of Nitz et al. (1999) confirms the accuracy of the
Cardon et al. scale:
log~$gf_{\rm CSSTW82} -$ log~$gf_{\rm NKWL99}$ = $-0.02$~dex
and $\sigma_{\rm CSSTW82}$ $\simeq$ 0.08~dex. We are thus confident that all
stars are on a uniform, accurate $gf$ scale. This is quite an achievement since
the stars span four orders of magnitude in abundance, and have been analysed by
several independent groups.

As the DLA Co abundances are derived from ionised lines whereas the stellar
transitions are neutral, there is (in principle) a possibility that the stellar
and DLA abundance scales differ. However, as the $gf$ scales for both 
ionisation
states are tied to modern lifetime measurements, we doubt that differences
exceed 5-10\%, or 0.02 -- 0.04~dex.

Co~I lines have extensive hyperfine structure which can affect abundance 
measurements by up to several $\times 0.1$~dex (e.g. Ryan et al. 1996, Fig.~1).
The stellar studies cited have all accounted for hyperfine structure,
which was not always the case in earlier analyses.  Moreover, work by
Pickering and colleagues (1998 and priv.comm.) have confirmed that the Co~II
lines we have observed in the DLAs will not be broadened sufficiently
to affect our results.

Both Co and Fe are primarily singly ionised in the temperature range of the 
stars investigated, so the Co~I measurements are of a minority state.
However, because these elements have almost identical first ionisation 
potentials, uncertainties in their temperatures or surface 
gravities will affect the ionisation degree of Co~I just as much as for Fe~I.
Representative errors in [Co/Fe] --- see Ryan et al. 1996, Table~3 ---
appear to be $\simeq$0.05-0.10~dex (1$\sigma$) for good S/N ($\sim$100) data, 
but can be $\sim0.2$~dex for poorer data with S/N $\simeq$ 30. It is believed
that the lower S/N levels achieved in the first studies of the most metal-poor
stars may explain most of the spread in [Co/Fe] seen in those stars.
(See Norris et al. 2001 for much improved [Co/Fe] measurements in such 
objects.)

\subsection{Comparison with Relative Stellar Abundances}

The stellar data discussed in the previous section are plotted in Figure~3, as 
a function of [Fe/H]. As discussed in the Introduction, Galactic stellar 
[Co/Fe] ratios show distinct --- but non-monotonic --- trends as a function of 
metallicity, which indicate key differences between the various stellar 
populations.  The halo and bulge data, for example, at high and low 
metallicity,
share great age and the highest [Co/Fe] ratios, whereas later-forming stars in
the disk have progressively lower [Co/Fe] ratios. 
In addition, the Co detection in Q2206$-$199 and
three upper limits are also plotted, although for the DLAs [Zn/H]
is used as a metallicity indicator rather than [Fe/H].  
The first determination of [Co/Fe] = $+0.31\pm0.05$ in a DLA is consistent
with the overabundance seen in Galactic halo stars
and the transition regime between the thick disk and bulge at
a metallicity consistent with the latter populations.  The point
plotted for Q2206$-$199 is the value determined from our Voigt
profile fit although, as discussed in section 4, this may be a
small overestimate if there is some blending of the blue components.  
In addition, Co is also overabundant relative to Cr in Q2206$-$199,
[Co/Cr] = +0.23, again consistent with thick disk, moderately
metal poor halo and metal-rich bulge stars.

\begin{figure}
\epsfxsize=084mm
\epsfbox[20 366 565 742]{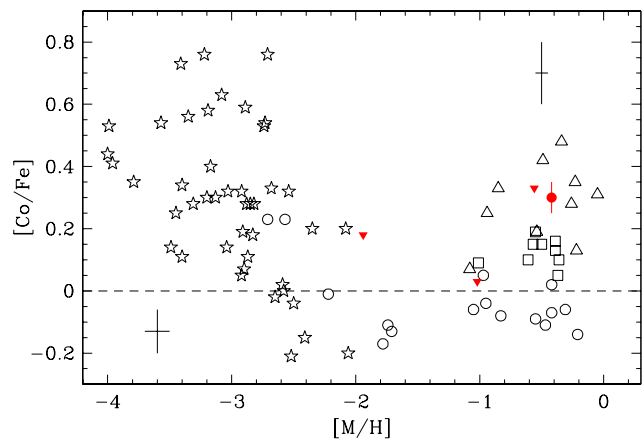}
\caption{Relative cobalt abundances as a function of metallicity for 
different Galactic populations and DLAs: 
open stars --- metal-poor halo stars from McWilliam et al (1995) and Ryan, 
Norris, \& Beers (1996);
open circles --- moderately metal-poor thin disk and halo stars from Sneden \& 
Gratton (1991), with $gf$ changes as described in the text;
open triangles --- bulge stars from McWilliam \& Rich (1994);
open squares --- thick-disk stars from Prochaska et al. (2000).
The solid circle represents Q2206-199 (with associated fitting error)
and the down-turned filled triangles are 
upper limits for three other DLAs.  The cross in the lower left corner
of the figure shows the typical error bar for stellar points, whereas
the error bar in the top right corner shows the maximum effect of systematics
(continuum placement and $N$(H~I) determination) discussed in the
text.}
\end{figure}

The trend of high [Co/Fe] in metal-poor stars was first pointed out by 
McWilliam et al. (1995) and confirmed by Ryan et al. (1996). 
The observation that stellar 
[Co/Fe] abundances are highest in the oldest populations might suggest that 
elevated abundances arise from the fastest evolving stars, i.e. the most 
massive supernova (SN) progenitors. However, halo star [Co/Fe] values are not 
uniformly high, and clearly vary as a function of metallicity. McWilliam et al.
(1995) suspected that the trends were due to a metallicity dependence of the 
yields of elements produced in supernova (SN) nucleosynthesis, while Ryan et 
al. (1996) noted that differences in the explosion energy could affect the 
yields in this way. It is unclear whether the metallicity-dependence is 
secondary, being driven by lower Co/Fe yields from later-evolving, lower mass 
SN, or whether it reflects an explicit dependence on progenitor metallicity. 
More than one factor may be (and probably is) at work.  Umeda et
al. (2000) have shown that the abundances of these elements do not depend
strongly on the initial stellar metallicity. Nakamura et al. (1999)
have shown that the abundance trends of Cr, Co and Mn observed in the most
metal-poor stars can be reproduced when changing the position of the mass
cut, but at the expense of overproducing Ni,
which is only rarely observed in the stars (Ryan et al. 1996).  
More recently, Nakamura et al. (2000) have also been able
to reproduce the trends, though not the absolute values, in Co and 
Cr by using nucleosynthesis
models of hypernovae, i.e. supernovae with very high explosion
energies.  These authors remark that as well as re-producing 
Galactic abundances, they can also explain the unusual patterns
seen in some starburst galaxies, such as M82.  Indeed,
several other observational clues indicate that star formation
in DLAs proceeds in short bursts separated by extended quiescent periods,
such as low N/O and [S,Si/Zn] values (Pettini et al. 2001;
Pettini et al 2000) and the lack of detected star formation through
H$\alpha$ imaging down to levels of $\sim 1$ M$_{\odot}$ yr$^{-1}$
(Bouche et al. 2000; Kulkarni et al. 2001).  However, both the models
of Nakamura et al. (1999; 2000) also predict an overabundance of Ni
relative to Fe, something that is rarely seen in stars, but quite commonly
seen in DLAs (Figure 4).  In fact all 4 of the DLAs plotted in Figure
3 exhibit [Ni/Fe] $\sim$ 0.2  \footnote{We note that Prochaska \& Wolfe 
(1997) use the old NiII f-values, requiring a 
correction by a factor of 1.9, which brings their N(Ni) into very good 
agreement with ours.}.  However, since at least HE1104$-$1805
shows no strong overabundance of Co, it appears that 
neither stars nor DLAs exhibit consistent or comprehended 
relative abundances of Co and Ni.  Finally, we note that since the revision
of Ni~II $f$-values, a systematic overabundance is now seen for [Ni/Fe]
in DLAs, perhaps indicating that further refinement of the atomic
data may still be required.

As detailed supernova nucleosynthesis models have yet to provide a convincing 
explanation of the [Co/Fe] abundances of stars, we would not presume to be 
able to solve that mystery with the first ever DLA observations. Nevertheless,
by comparing the DLA data with the stellar measurements, useful insights 
can be 
obtained. As noted, the [Co/Fe] ratio in the $z_{\rm abs} = 1.92$ DLA of 
Q2202$-$199 falls near the boundary between bulge and thick disk stars. It is 
seen at a lookback time of $\simeq$8.5~Gyr, around the time of (or a few Gyr 
after) the formation of the oldest thick disk and bulge objects. However, 
the possible 
absence of an elevated [Co/Fe] ratio in Q1223+17, with a lookback time 
slightly ($\simeq0.5$~Gyr) longer, emphasises that age alone does not explain 
the elevated [Co/Fe] values. If massive stars do produce the elevated 
[Co/Fe] ratios, then possibly star formation started later in the $z = 1.92$ 
DLA than in the others.  This postulate could be further explored by
considering the abundance of Si, an $\alpha$ element.  We find 
[Si/Zn] =$-0.07$, although this value is likely be be an underestimate
of the true Si abundance due to dust depletion.  It is possible
to make a crude adjustment for the depletion fraction by considering 
Zn and Cr ratios, which are found to be solar in stars.  Therefore,
any significant departure from [Zn/Cr] = 0 can be explained by
the incorporation of Cr into grains.  For
Q2206$-$199, we determine [Zn/Cr] = 0.35 dex, applying this correction
to Si gives [Si/Zn] $\sim$ 0.3, although this is likely to be an
upper limit since Si is less depleted than Cr in the local ISM
(Savage and Sembach 1996).  In any case, it is likely that Si
exhibits at least a mild overabundance relative to Zn, consistent
with the expectation if the enrichment of this galaxy's ISM has
been dominated by the products of Type II SN. 

Nakamura et al. (1999, 2000) have achieved partial success in attempting to 
explain the iron-group element ratios (Ni excepted) in their nucleosynthesis 
models, firstly by moving the mass-cut and secondly by changing the explosion 
energy. The latter, as they note, partially simulates the former effect because
the zone of complete Si-burning is shifted outward in high-energy models 
compared to classical 10$^{51}$~erg models. However, increasing the energy has 
the additional beneficial outcome that it provides a solution, {\it for the 
first time}, to the long-standing problem of explaining why Ti is overabundant 
in halo stars.  It is found that the greater radial extent over 
which Si-burning occurs in the high-energy models necessarily 
incorporates lower density zones, which allows for increased 
$\alpha$-rich freeze-out-production of $^{48}$Ti.  Previous classical 
SN nucleosynthesis models always yielded [Ti/Fe] $\simeq$ 0.
If this finding may be considered independent evidence that higher-energy 
explosions are the norm in the more massive SN, i.e. those that enriched the 
extremely metal-poor halo and the `get-rich-quick' bulge, then we should 
understand the higher [Co/Fe] and simultaneously the higher [Ti/Fe] ratios in 
these populations that were enriched by massive-stars.  However,
Prochaska \& Wolfe (1997) find that [Ti/Fe] = $-0.07$ in Q2206$-$199, 
although considering that Ti is slightly more depleted than 
Fe (Savage and Sembach 1996), this DLA is likely to have a 
roughly solar Ti/Fe ratio.

Once high-energy SN models are explored more widely and included into
galactic chemical evolution calculations, and a larger range of elements 
including Co and possibly Ti have been measured in more DLAs, 
the complementary 
constraints presented by DLA and stellar systems can be considered jointly. 
Nevertheless, it is important to appreciate through all of this that the 
similarities of DLA [Co/Fe] and [Co/Cr] ratios to those of a given Galactic 
population do not necessarily imply that this DLA, or the DLA population in 
general, represents the precursor to modern day halos/bulges. Rather, it is
indicative of similar physical properties such as star formation rate and mix 
of SN types between the populations.

\begin{figure}
\epsfxsize=084mm
\epsfbox[20 366 583 743]{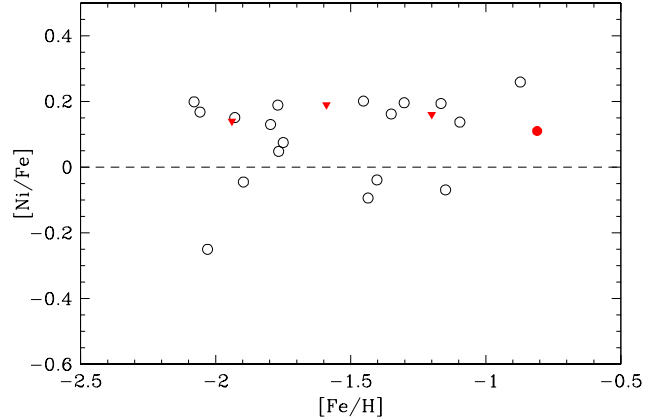}
\caption{Ni and Fe abundances in DLAs.  Open circles are values taken
from the literature (Prochaska et al 2001; Pettini et al. 2000),
filled triangles are the DLAs for which a Co upper limit has been
obtained herein and the filled circle is for the DLA towards
Q2206$-$19 for which Co has been detected.}
\end{figure}

\section{Summary And Conclusions}

The distinct trends exhibited by Co in the various Galactic
stellar populations makes this an interesting element to target
in DLA abundance studies.  However, the weakness of the Co~II
lines renders this a challenging prospect, even for an 8-m
class telescope such as the VLT.  Nonetheless, we have shown
that by selecting DLAs with high column densities of Fe towards
bright QSOs, the detection of Co is feasible with instruments
such as UVES and HIRES down to [Co/Fe] $\sim$ 0.  We have
achieved the first detection of [Co/H] = $-0.50$ 
in one of the DLAs studied herein
and a sensitive upper limit for a second system.  In addition, we
have determined useful upper limits for [Co/H] for two other DLAs
from the literature.  The detection of Co towards Q2206$-$199
shows an overabundance relative to Fe, [Co/Fe] = +0.31$\pm0.05$, 
at a metallicity
of approximately $\frac{1}{3} Z_{\odot}$, consistent with both the
ratios observed in Galactic bulge and thick disk stars.  
In fact, Prochaska et al (2000) noted that there is evidence
from Galactic bulge/thick disk stellar
abundances that these two populations may
have shared a similar formation history.  One current limitation
of the comparison with bulge stars, however, is that the
only measurements  presently available in
the literature were obtained with a 4-m telescope with relatively
low ($R \sim 17,000$) resolution (McWilliam and Rich 1994).
Repeating these observations with Keck/HIRES has shown that these
abundances may undergo significant revision, for example [Fe/H]
has already been found to be higher by $\sim$ 0.2 compared with
the 4-m data of lower resolution and S/N (Rich and McWilliam 2000).
The revised Co abundances from the HIRES data are not yet available,
but will eventually provide an interesting comparison for our DLA
detection.  Conversely, the upper limit
of [Co/Fe] $\le$ +0.18 that we determine for Q1223+17 is not consistent 
with the bulge star pattern and is more in line with Galactic disk
abundances, as is the result for HE1104$-$1805.

However, the interpretation of the data presented here are 
limited by statistics
and we clearly can not presume to understand the star formation history
of DLAs based on a single detection.  In addition, Q2206$-$199
is already known to exhibit large abundances of other elements
which are not normally seen in DLAs, such as Ti (Prochaska
and Wolfe 1997).  We can therefore not exclude the possibility that
Q2206$-$199 exhibits peculiar abundances that may not be seen
in other such systems.  Finally, we have deliberately chosen
DLAs with high $N$(Fe) for this initial study in order to maximise
the probability of a Co detection.  In order to extend this analysis,
such a selection bias will eventually have to be overcome if we
are to truly understand the trends of relative abundances.  Nonetheless,
this is an interesting new result and the study of Co in DLAs clearly 
invites further study.

Our understanding of the processes that drive abundance trends in
stars has only truly become possible thanks to large datasets,
since intrinsic scatter and errors will dominate small samples.
Therefore, more DLA data points are clearly required in order
to draw firm conclusions about the star formation histories implied
by the observed abundance trends.  Perhaps most importantly, we have shown
that with current 8-m class telescopes and high resolution echelle
spectrographs that such measurements are not only possible, but also
that Co may be a rewarding element to study in the context of
testing nuclesynthetic models and understanding galactic star
formation histories.

\section*{Acknowledgments}

We are very grateful to our collaborators Max Pettini, Jacqueline 
Bergeron and Patrick Petitjean for consenting to release the data for
Q2206$-$199, originally obtained for a different project.
It is a pleasure to acknowledge the staff of the VLT and Keck for 
their assistance during the acquisition of these data.  In particular
we are grateful to Andreas Kaufer for his support during both
the VLT observations and data reduction.
We would like to thank Chris Howk and Emmanuel Jehin for helpful 
discussions and to Max Pettini and Norbert Christlieb
for useful comments on an earlier draft.  We are
grateful to Sebastian Lopez for providing his Co upper limit in the
DLA towards HE1104$-$1805.

\label{lastpage}


\begin{thebibliography}{}



\bibitem[\protect\citename{Ballester et al.} 2000]{bal00}%
	Ballester, P., Modigliani, A., Boitquin, O., Cristiani, S.,
	Hanuschik, R., Kaufer, A., Wolf, S., 2000, ESO Messenger,
	101, 31

\bibitem[\protect\citename{Bouche et al.} 2000]{bou00}%
	Bouche, N., Lowenthal, J., Charlton, C., Bershady, M., 
	Churchill, C., Steidel, C., 2000, ApJ, accepted, astro-ph/0011374


\bibitem[\protect\citename{Cardon et al} 1982]{car82}%
	Cardon, B. L., Smith, P. L., Scalo, J. M., Testerman, L., 
	\& Whaling, W.  1982, ApJ, 260, 395



\bibitem[\protect\citename{Centurion et al.} 2000]{cen00}%
	Centuri\'{o}n, M., Bonifacio, P., Molaro, P., \& Vladilo, G.
	2000, ApJ, 536, 540

\bibitem[\protect\citename{Chen et al.} 1999]{chen99}%
        Chen, Y.Q., Nissen,P.E., Zhao, G., Zhang, H.W., \&
        Benoni, T. 2000, A\&AS, 141, 491




\bibitem[\protect\citename{Ellison et al.} 2000]{sara00}%
	Ellison, S. L., Yan, L., Hook, I., Pettini, M., Shaver, P.,
	Wall, J., 2000, ESO Messenger, 102

\bibitem[\protect\citename{Fedchak and Lawler} 1999]{fl99}%
	Fedchak, J. A., \& Lawler, J. E.
	1999, ApJ, 523, 734

\bibitem[\protect\citename{Federman et al.} 1993]{fed93}%
	Federman, S., Sheffer, Y., Lambert, D., Gilliland, R.,
	1993, ApJl, 413, 51




\bibitem[\protect\citename{Gratton and Sneden} 1991]{gs91}%
	Gratton, R., \& Sneden, C., 1991, A\&A, 241, 541







\bibitem[\protect\citename{Howk and Sembach} 1999]{hs99}%
        Howk, J.C., \& Sembach, K.R. 1999, ApJ, 523, L141


\bibitem[\protect\citename{Izotov, Schaerer, Charbonnel} 2000]{isc00}%
	Izotov, Y., Schaerer, D., Charbonnel, C., 2000, ApJ,
	accepted, astro-ph/0010643

\bibitem[\protect\citename{Izotov and Thuan} 1999]{it99}%
	Izotov, Y. I. \& Thuan, T. X., 1999, ApJ, 511, 639






\bibitem[\protect\citename{Kobulnicky and Zaritsky} 1999]{kz99}%
        Kobulnicky, H.A., \& Zaritsky, D. 1999, ApJ, 511, 118

\bibitem[\protect\citename{Kulkarni et al.} 2000]{kul00}%
	Kulkarni, V., Hill, J., Schneider, G., Weymann, R.,
	Storrie-Lombardi, L., Rieke, M., Thompson, R., Jannuzzi, B.,
	2000, ApJ, accepted, astro-ph/0012140


\bibitem[\protect\citename{Lattanzio et al.} 2001]{lat01}%
	Lattanzio, J., Pettini, M., Tout, C. A., \& Carigi, L.
	2000, A\&A, submitted



\bibitem[\protect\citename{Levshakov, Kegel, Agafonova} 2000]{lka00}%
	Levshakov, S., Kegel, W., Agafonova, I., 2000, A\&A, submitted
	astro-ph/0011513

\bibitem[\protect\citename{Lopez et al} 1999]{seb99}%
	Lopez, S., Reimers, D., Rauch, M., Sargent, W., Smette, A.,
	1999, ApJ, 513, 598

\bibitem[\protect\citename{Lu, Sargent and Barlow} 1998]{lsb98}%
	Lu, L., Sargent, W. L. W., Barlow, T.A.
	1998, AJ, 115, 55




\bibitem[\protect\citename{McWilliam and Rich} 1994]{mr94}%
	McWilliam, A., \& Rich, M., 1994, ApJS, 91, 749

\bibitem[\protect\citename{McWilliam et al.} 1995]{mpss95}%
	McWilliam, A., Preston, G. W., Sneden, C., Searle, L.,
	1995, AJ, 109, 2757



\bibitem[\protect\citename{Mullman et al.} 1998a]{mul98a}%
	Mullman, K., Cooper, J., Lawler,  1998,
	ApJ, 495, 503

\bibitem[\protect\citename{Mullman et al.} 1998b]{mul98b}%
	Mullman, K., Lawler, J., Zsargo, J., Federman, S., 1998,
	ApJ, 500, 1064

\bibitem[\protect\citename{Nakamura et al.} 2000]{nak00}%
	Nakamura, T., Umeda, H., Iwamoto, K., Nomoto, K., Hashimoto, M.,
	Hix, R., Thielemann, F., 2000, ApJ, submitted, astro-ph/0011184

\bibitem[\protect\citename{Nakamura et al.} 1999]{nak99}%
	Nakamura, T., Umeda, H.,  Nomoto, K., Thielemann, F., Burrows, A.,
	1999, ApJ, 517, 193

\bibitem[\protect\citename{Nitz et al} 1999]{nitz99}%
	Nitz, D. E., Kunau, A. E., Wilson, K. L., \& Lentz, L. R. 
	1999, ApJS, 122, 557


\bibitem[\protect\citename{Norris, Ryan \& Beers} 2001]{nrb01}%
	Norris, J. E., Ryan, S. G., \& Beers, T. C. 2001, in prep.
 


\bibitem[\protect\citename{Pettini et al.} 1999]{pet99}%
	Pettini, M., Ellison, S. L., Steidel, C. C., Bowen, D. V.
	1999, ApJ, 510, 576

\bibitem[\protect\citename{Pettini et al.} 2000]{pet00}%
	Pettini, M., Ellison, S. L., Steidel, C. C., 
        Shapley, A. E., \& Bowen, D. V.
	2000, ApJ, 532, 65

\bibitem[\protect\citename{Pettini et al.} 2001]{pet01}%
	Pettini, M., et. al
	2001, in preparation



\bibitem[\protect\citename{Pettini et al} 1997]{pet97}%
        Pettini, M.,  Smith, L.J., King, D.L., \& Hunstead, R.W. 
        1997, ApJ, 486, 665

\bibitem[\protect\citename{Pettini et al.} 1994]{pet94}%
        Pettini, M.,  Smith, L.J., Hunstead, R.W., King, D.L., 
        1994, ApJ, 426, 79

\bibitem[\protect\citename{Pickering et al} 1998]{pick98}%
	Pickering, J. C., Raassen, A. J. J., Uylings, P. H. M., \& 
	Johansson, S. 1998, ApJS, 117, 261

\bibitem[\protect\citename{Prantzos and Boissier} 2000]{pb00}%
	Prantzos, N., Boissier, S., 2000, MNRAS, 315, 82

  
\bibitem[\protect\citename{Prochaska et al.} 2001]{x01}
	Prochaska, J. X., et al., 2001, in preparation.

\bibitem[\protect\citename{Prochaska et al.} 2000]{x00}
        Prochaska, J. X., Naumov, S.O., Carney, B.W., McWilliam,
        A., \& Wolfe, A. M. 2000, ApJ, accepted


\bibitem[\protect\citename{Prochaska and Wolfe} 1997]{pw97}%
	Prochaska, J. X., \& Wolfe, A. M.
	1997, ApJ, 474, 140


\bibitem[\protect\citename{Prochaska and Wolfe} 1999]{pw99}%
	Prochaska, J.X., \& Wolfe, A.M.
	1999, ApJS, 121, 369





\bibitem[\protect\citename{Rich and McWilliam} 2000]{rm00}%
	Rich, M., \& McWilliam, A., 2000, Proceedings of the SPIE vol. 4005,
	astro-ph/0005113 

\bibitem[\protect\citename{Ryan, Norris, \& Beers} 1996]{1996}%
	Ryan, S. G., Norris, J. E., \& Beers, T. C., 1996, ApJ, 471, 254

\bibitem[\protect\citename{Savage and Sembach} 1996]{ss96}%
	Savage, B. D. \& Sembach, K. R., 1996, ARA\&A, 34, 279

\bibitem[\protect\citename{Savage and Sembach} 1991]{ss91}%
	Savage, B. D. \& Sembach, K. R., 1991, ApJ, 379, 245

\bibitem[\protect\citename{Savaglio} 2000]{sav00}%
	Savaglio, S., 2000, Invited talk at the IAU Symposium 204 
	`The Extragalactic Infrared Background and its Cosmological
    	Implications', Manchester, August 2000, eds. M. Harwit \& M.G. Hauser,
	astro-ph/0011473


\bibitem[\protect\citename{Shetrone, Cote and Sargent} 2000]{scs00}%
	Shetrone, M., Cote, P., Sargent, W. L. W., 2000, ApJ 
	accepted, astro-ph/0009505









\bibitem[\protect\citename{Tinsley} 1979]{t79}%
	Tinsley, B. M. 1979, ApJ, 229, 1046


\bibitem[\protect\citename{Umeda et al.} 2000]{um00}%
  	Umeda, H., Nomoto, K., Nakamura, T., 2000, in The First
	Stars, pp. 150--174 (astro-ph/9912248)

\bibitem[\protect\citename{Viegas} 1995]{v95}%
        Viegas, S.M. 1995, MNRAS, 276, 268

\bibitem[\protect\citename{Vladilo} 1998]{v98}%
	Vladilo, G., 1998, ApJ, 493, 583








\end{thebibliography}
\end{document}